\documentclass[letterpaper,12pt]{article}

\usepackage{graphicx}

\title{On the Behavior of Journal Impact Factor Rank-Order Distribution}
\author{Mansilla, R.(1), K\"oppen, E.(2) Cocho, G.(2) and Miramontes, P.(3,*)\\
1. Centro de Investigaciones
Interdisciplinarias. Universidad
Nacional \\
Aut\'onoma de M\'exico. M\'exico 04510, DF. M\'exico\\
2. Instituto de F\'isica. 
Universidad Nacional Aut\'onoma de M\'exico\\
M\'exico 04510, DF. M\'exico\\
3. Facultad de Ciencias. Universidad Nacional Aut\'onoma de M\'exico\\
M\'exico 04510, DF. M\'exico\\ 
* Corresponding author. Email: pmv@fciencias.unam.mx}

\begin{document}

\maketitle

\begin{abstract} 

An empirical law for the rank-order behavior of journal impact factors is
found. Using an extensive data base on
impact factors including journals on Education, Agrosciences,
Geosciences,Mathematics, Chemistry, Medicine, Engineering, Physics, Biosciences
and Environmental, Computer and Material Sciences, we have found extremely good
fittings outperforming other rank-order models. Based in our results we propose
a two-exponent Lotkaian Informetrics. Some extensions to other areas of
knowledge are discussed.

\end{abstract}

\section{Introduction}

Quantitative studies in linguistics have a long lineage. Due
to the extreme complexity of languages, these studies have been mainly based on
statistical
properties of words in literary corpora. Outstanding early examples of these
studies are J. B. Estoup (1916), G. Dewey (1923) and E. V. Condon (1928).
However the most influential contribution on this topic is by G. K. Zipf (1949).
In his work appears what is today known as \textit{Zipf's law} which can be
formulated as follows: Let $f(r)$; $r=1,\ldots,N$,  be the relative frequency
of the words in a text in decreasing order. Then Zipf's law states that:

\begin{equation}
f(r)=\frac{K}{r^\alpha}
\end{equation}

In this case, the items are words taken from a given text, the
most abundant
word takes the first place ($r=1$), the second one takes the following place
($r=2$) and so on. The fact that the mathematical expression of the law is a
negative exponent power law implies that the law is a straight line with
negative slope $\alpha$ when plotted in log-log scales. $K$ is a
proportionality constant with no phenomenological interest.
This empirical law has found
applications in a wide range of natural and human phenomena (Li, 2003). The case
when $\alpha \simeq 1$ is of particular
interest because it implies self-similarity . The exact mechanism behind
Zipf's law still remains a mystery so far. However it is important to remark
that the presence of power laws implies in general that the underlying
mechanism is neither stochastic or regular. Power laws are the signature of
correlated noise possibly associated to and "edge of chaos" dynamics (REF) or
could be a clue to self-organized criticality (Bak et al, 1989)

The main drawback of Zipf's law was the bad fitting at very high and very low
frequencies in the word counting problem. An improvement over the
Zipf's law was proposed by B. Mandelbrot (1954):

\begin{equation}
f(r)=\left[ \frac{N+\rho}{r+\rho}\right]^{1+\epsilon}
\end{equation}

Where  $N$ is the number of different words in the text and $\rho$, $\epsilon$ 
are parameters to be adjusted. 

Zip's law is a special case of Mandelbrot's. This fact, along with a complete
discussion of the role of power laws in the field of Informetrics can be found
in (Egghe, 2005). 

Recently it has been reported (Le Quan; 2002) that what Zipf found is valid for
small corpora (for the size of the text that were analyzable at that time), and
that today that the computer allows the analysis of huge texts, the log-log plot
shows a clear downwards bending tail instead of the predicted straight line.

Scientific productivity is another topic where the first studies date back
almost a century with the works of A. Dresden (1922) and A. J. Lotka (1926).
The law of Lotka has the same mathematical form of eq (1) but he already
introduced bibliometric variables by using $r$ as contributors or authors of 
a given paper and $f(r)$ as articles or papers themselves. Since Lotka, it is
common to call ``sources" the independent variable and ``item" the dependent
one. This way, Lotka's law states that the number of items is a power law of the
sources. The branch of Informetrics related to the study of power laws is
called \textit{Lotkaian Informetrics} (Egghe, 2005).

Informetrics mainly deals with the relationships between sources and items.
It is normal to find the pairs authors-journals or journals-bibliographies as
sources and items. In this paper we explore the possibility of extending the
Lotkaian Informetrics to the realm of Journal Impact Factors (JIFs). We show as
well that the rank-order JIFs plots deviate from a traditional Lotkaian
equation and
propose an extension to what it could be called \textit{two-exponent Lotkaian
laws}.

\section{Impact Factors}

Impact Factor is a measure of the frequency with which the "average article" in
a journal has been cited in a particular year or period (Garfield; 1994), it
is calculated "by dividing the number of times a journal has been cited by
the number of articles it has published during some specific period of time.
The journal impact factor will thus reflect an average citation rate per
published article" (Garfield; 1955). The impact factor of journals is an attempt
to evaluate the knowledge production published among different journals of a
given field. Mainly
covered by the\textit{ Science Citation Index} database, it is published
annually since 1975 in the \textit{Journal Citation Reports}.

JIFs has been the target of many criticisms (Soegler; 1997, Fr\"ohlich, 1996)
and there is a
debate about its usage as a tool to evaluate research. Even the
influential journal \textit{Nature} states that the JIFs figures should
be handled carefully (Nature; 2005). Regardless its pros and
cons, the fact is that it is an every day measure of the importance of a
journal and it is worlwide used (de Marchi and Rocchi; 2001). 

While keeping a skeptical attitude towards the use of the JIFs to evaluate
scientific research, it should be recognized that it is an outcome of the
process of publication and it has became by itself a subject of
scientific study. 

Rank-order distribution of JIFs attracted the attention of D. Lavalette who
(mentioned in Popescu (2003) proposed the following law:

\begin{equation}
f(r)=K \left[ \frac{N+1-r}{r}\right]^b
\end{equation}

Where $N$ is the number of journals, $r$ is the ranking number, $f(r)$
is the impact factor, $b$ is a parameter to be fitted.

In the next section we propose a law that outperforms Lavalette's (see
Concluding Remarks).

\section{Analytical expression of the law}

Figure 1 shows the log-log plot of the IF of a randomly taken field from
Popescu's database (2003)

\bigskip
\begin{center}
Figure 1
\end{center}
\bigskip

It is evident that it is not a power law because of the
bending tail in the right side of the plot. This fact motivated us to propose a
Beta-like function:

\begin{equation}
f(r)=K \frac{(N+1-r)^b}{r^a}
\end{equation}

$f(r)$, $r=1 \ldots,N$ represents the rank-order impact factors; $K$,
$a$ and $b$ are three parameters to fit. $K$ is a meaningless scaling factor.
Notice that when $b=0$ this equation becomes Lotka's law.

\section{Results}

For every set of data, we find the
parameters values using a linear least squares method on the logarithmic
variable:

\begin{equation}
log(f(r))=log(K)+blog(N+1-r)-alog(r)
\end{equation}

Table 1 shows the values of $K$, $b$ and $a$, as well as the coefficient of
regression $r^2$ for impact factors of twelve disciplines.  In Figs 2, 3 and 4
the
impact factors data as well as our theoretical curve for the fields of Physical,
Mathematical and Environmental Sciences are shown. We used semilog plots
because they are more natural when the abscissa is a rank-order variable.

\bigskip
\begin{center}
Table 1.
\end{center}
\bigskip
\begin{center}
Figures 2, 3 and 4.
\end{center}

\section{Concluding remarks}

We have shown the excellent agreement of the data with our model. The quality of
the fitting is superior to the proposal of Lavalette. From the
comparison of eqs 3 and 4, it is evident that Lavalette's law is a
particular case of ours when $a=b$. Unfortunately, it is not possible to
discuss the rationale behind Lavalette's law because the original paper is not
available and all we know about it is a mention in Popescu's
paper (2003).

The underlying proposed mechanism yielding the above discussed behaviors often
assumes a kind of ``biological evolution form". For instance, G. Yule (1924)
working in a model suggested by J. Willis (1922) managed to prove that assuming
a single ancestral specie and probabilities of mutation and duplication a power
law
behavior is obtained. Expansion-modification systems proposed by W. Li (1991),
which take into account the basic features of DNA mutation processes (R.
Mansilla and Cocho, 2000), are also able to predict this behaviour.

When discussing journal impact factors, a balance between the importance to the
researchers of publish their work in high ranked journal, the difficulties
associated with doing this and the increase of impact received by journals with
high impact factor, seems to create a ``rich gets richer" (the "Matthew
Effect", see (Merton; 1968 and Egghe and Rousseau; 1999) mechanism also
observed in complex networks (Barabasi, 2002). More than 49 years ago, H. Simon
(1955, 1957) proposed a model which produces similar distributions. It is also
interesting to notice that the bending of the tail of JIFs rank-order
distribution means that after a critical zone of JIFs values is smooth thus
discarding the possibility of the existence of multifractality.

Power-laws seem to be ubiquitous in physics, biology, geography, economics,
linguistics, etcetera (see Li, W., 2003). We consider ``linguistic studies" not
only those related with natural languages but also arbitrary languages over
abstract finite alphabets. When the number of possible ``words" is large, as it
is the case for natural languages, it is expected to have a good fitness with a
one-parameter power law. However, when the number of words is rather
small, as it is the case of programing languages, one-exponent power
laws absolutely fails and more parameters are necessary for a suitable fit. New
elements to this considerations have been given by LeQuan et al (2002). They
showed that there is a serious deviation when the size of the sample is huge.

We expect that the increase in computing power will show that the deviation
of Zipf's and Lotka's laws is a generic phenomenon. Then, a two-exponent
Lotkaian and Zipfian infometrics and linguistics should be welcome. 

\section{Acknowledgements}

This work has been partially supported by the UNAM-PAPIIT grant IN-111003. The
authors thank the sound comments of two anonymous referees.

\newpage
\section{Table 1 }

\begin{tabular}{cccccc}
Scientific Field& & $k$& $b$& $a$& $R^2$ \\ 
-----& &  -----& -----& -----& ----- \\ 
Physics& & 0.0273 & 0.991& 0.4058& 0.9999 \\ 
Mathematics& & 0.0437 & 0.676& 0.2622& 0.9999 \\
Computer Science& & 0.0066 & 1.0626& 0.2840& 0.9999 \\ 
Agroscience& & 0.0070 & 0.9597& 0.2210& 0.9999 \\
Environmental Sc.& & 0.0358& 0.9357& 0.2781& 0.9800 \\ 
Biosciences& & 0.0304 & 1.0161& 0.5140& 0.9999 \\
Chemistry& & 0.0549 & 0.9733& 0.4560& 0.9999 \\ 
Engineering& & 0.0033 & 1.0472& 0.3522& 0.9999 \\
Geosciences& & 0.0463 & 0.8739& 0.3505& 0.9999 \\ 
Material Science& & 0.0408 & 0.9072& 0.4477& 0.9999 \\
Medicine& & 0.0819 & 0.7735& 0.4307& 0.9999 \\
Education& & 0.0819& 0.7735& 0.4307& 0.9999 \\

\end{tabular}

\newpage

\bigskip
\begin{figure}
 \centering
 \includegraphics[width=3.5in,bb=   81   227   529   564]{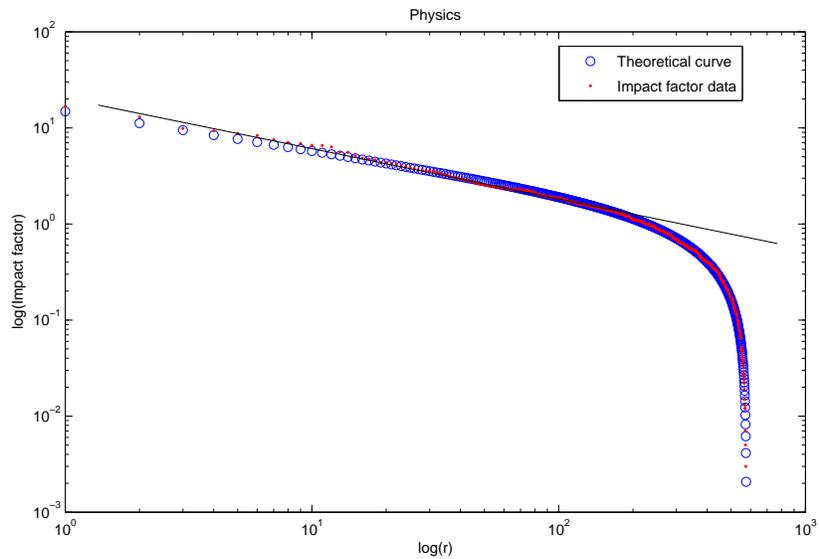}
 \caption{Log-log rank-order plot of the Impact Factor data for Physics
Journals. Notice the drop of the tail of the curve (see text)}
 \label{fig:1}
\end{figure}

\begin{figure}
 \centering
 \includegraphics[width=3.5in,bb=   81   227   529   564]{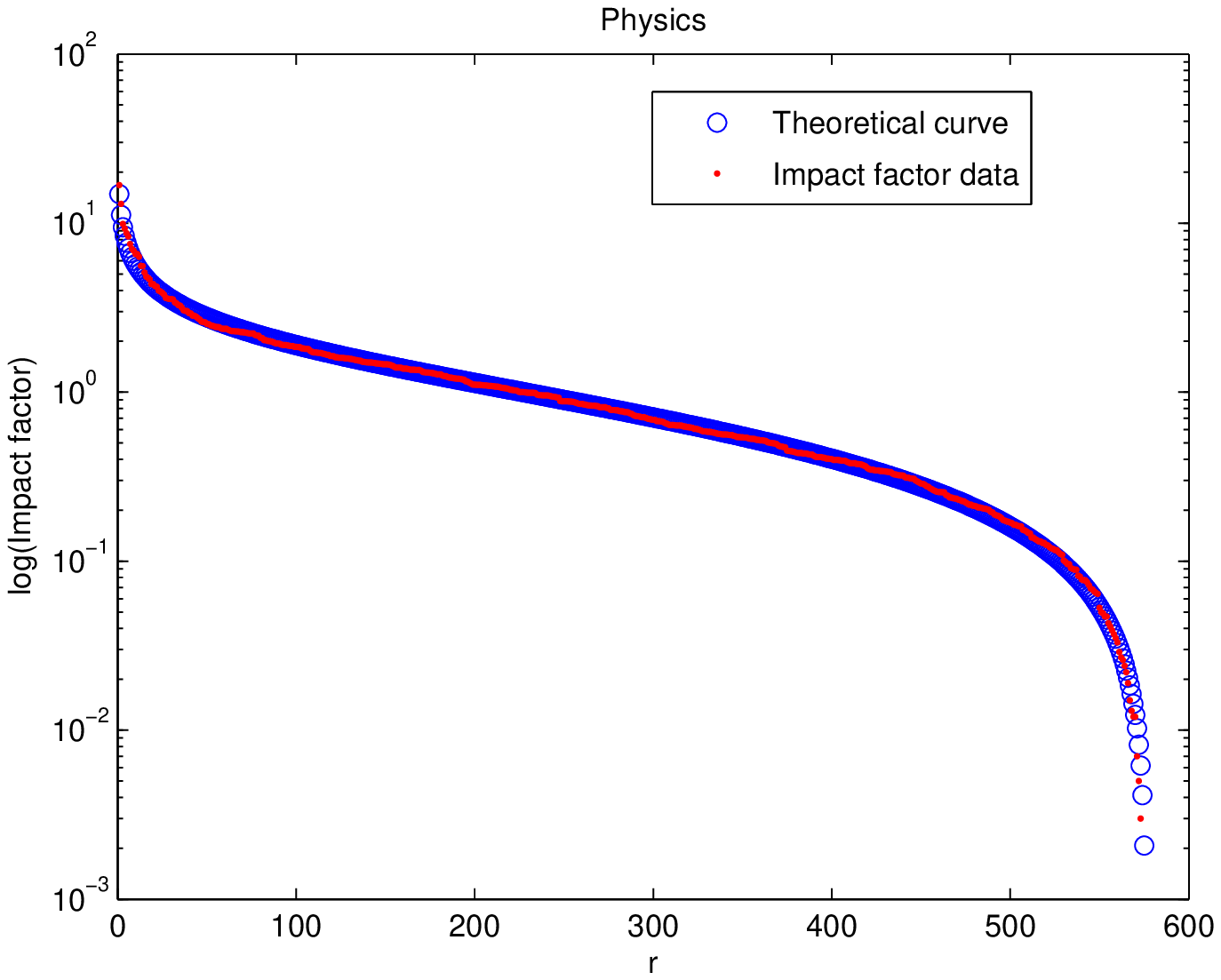}
 \caption{Semi-log Impact factor rank-order distribution for Physics journals.
Solid circles
represent raw data. Hollow circles are the ouput of the model}
 \label{fig:2}
\end{figure}

\begin{figure}
 \centering
 \includegraphics[width=3.5in, bb=   81   227   529   564]{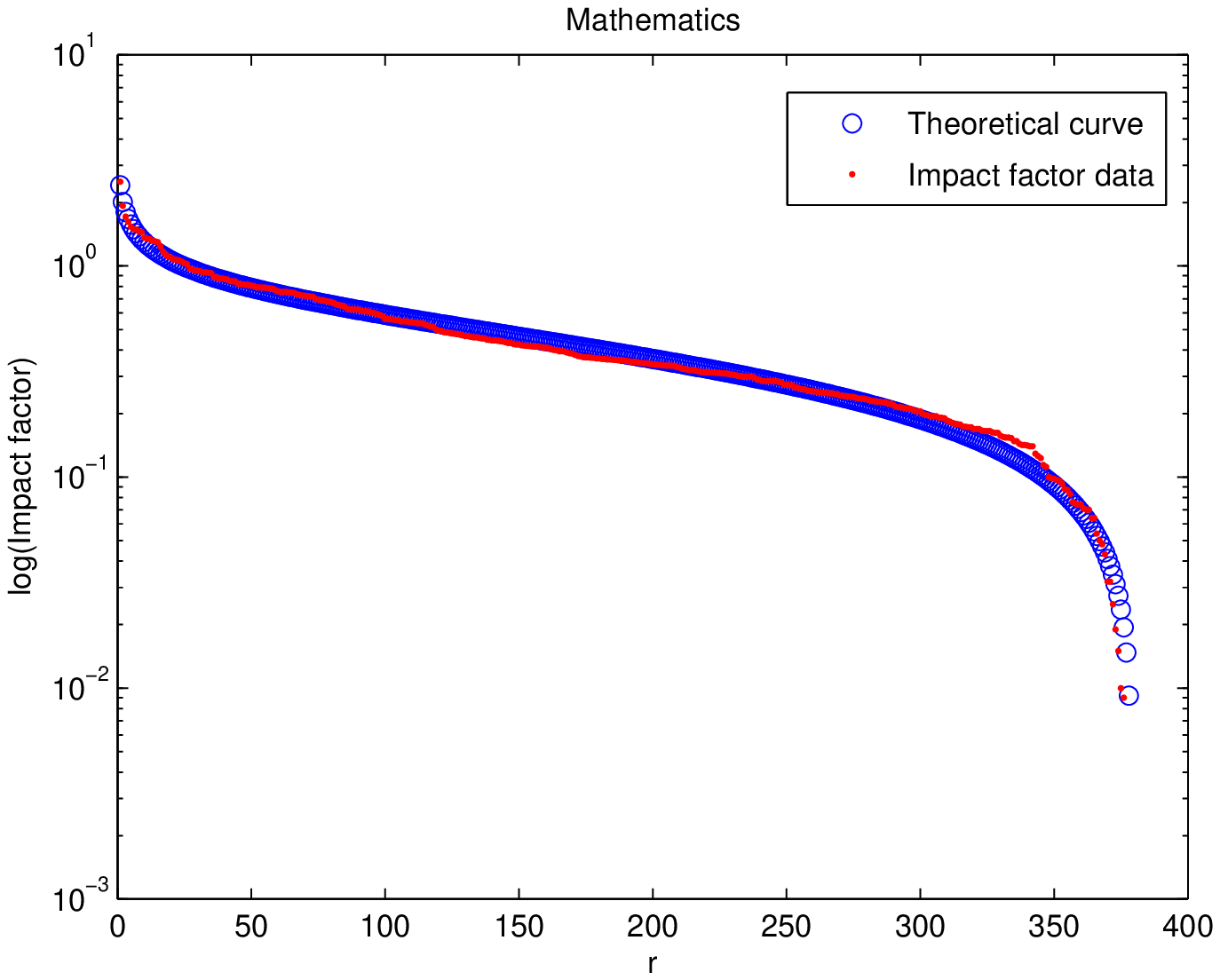}
 \caption{Semi-log Impact factor rank-order distribution for Mathematics
journals. Solid circles
represent raw data. Hollow circles are the ouput of the model}
 \label{fig:3}
\end{figure}

\begin{figure}
 \centering
 \includegraphics[width=3.5in,bb=   81   227   529   564 ]{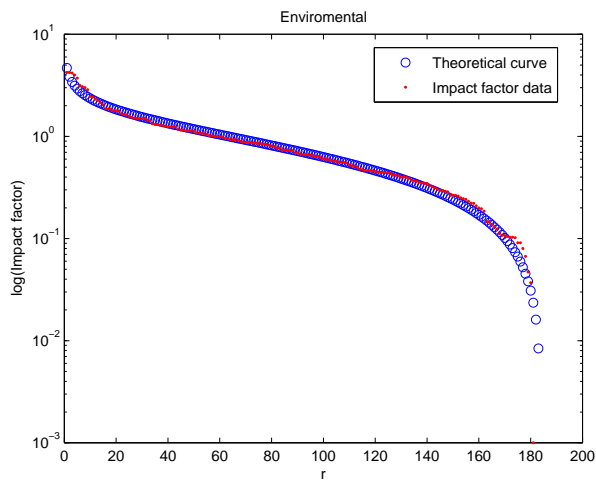}
 \caption{Semi-log Impact factor rank-order distribution for Environmental
sciences. Solid circles
represent raw data. Hollow circles are the ouput of the model}
 \label{fig:4}
\end{figure}

\end{document}